\documentclass[%
conference,
a4paper,
]{IEEEtran}
\usepackage{graphicx}
\graphicspath{{./../../../img/}}
\usepackage[noadjust]{cite}

\usepackage{ragged2e}
\usepackage{float}
\usepackage{graphicx}

\usepackage[english]{babel}
\usepackage[utf8x]{inputenc}
\usepackage{xcolor}
\usepackage{listings}
\usepackage{color}
\usepackage{lmodern}
\usepackage{tikz}
\usepackage{caption}
\usetikzlibrary{calc}
\usepackage{xparse}
\usepackage{booktabs}
\usepackage{dcolumn}
\usepackage{mathrsfs}
\usepackage{float}
\usepackage[absolute, overlay]{textpos}
\usepackage{hyperref}
\usepackage{nth}
\usepackage{amsmath}
\usepackage[linesnumbered,ruled]{algorithm2e}
\usepackage{color,soul}

\def\BibTeX{{\rm B\kern-.05em{\sc i\kern-.025em b}\kern-.08em
    T\kern-.1667em\lower.7ex\hbox{E}\kern-.125emX}}
    
\newcommand{\head}[1]{\textnormal{\textbf{#1}}}

\newcommand{\RR}{\mathbb{R}}

\usepackage{amsmath, amssymb}

\DeclareMathOperator{\simi}{sim}
\DeclareMathOperator{\loc}{loc}
\DeclareMathOperator{\glo}{glo}

\DeclareMathOperator{\feat}{feat}

\makeatletter
\newcommand*{\rom}[1]{\expandafter\@slowromancap\romannumeral #1@}
\makeatother

\usepackage{todonotes}

\urldef{\mails}\path|{firstname.lastname}@uni-passau.de|

\begin{document}


\title{Global and Local Feature Learning \\
  for Ego-Network Analysis}

\author{%
  \IEEEauthorblockN{%
    Fatemeh Salehi Rizi, Michael Granitzer,  Konstantin Ziegler
  }
  \IEEEauthorblockA{%
    Department of Computer Science and Mathematics\\
    University of Passau, Germany\\
    Email: \mails
  }
}

\maketitle


\begin{abstract}

  In an ego-network, an individual (ego) organizes its friends
  (alters) in different groups (social circles). This social network
  can be efficiently analyzed after learning representations of the
  ego and its alters in a low-dimensional, real vector space. These
  representations are then easily exploited via statistical models for
  tasks such as social circle detection and prediction. Recent
  advances in language modeling via deep learning have inspired new
  methods for learning network representations. These methods can
  capture the global structure of networks. In this paper, we evolve
  these techniques to also encode the local structure of
  neighborhoods. Therefore, our local representations capture network
  features that are hidden in the global representation of large
  networks. We show that the task of social circle prediction
  benefits from a combination of global and local features generated
  by our technique.

\end{abstract}

\begin{IEEEkeywords}
  Ego-Networks, Global Representations, Local Representations, Deep
  Learning, Graph Embeddings, Social Network Analysis
\end{IEEEkeywords}

\section{Introduction}

\begin{figure}
\centering
\includegraphics[width=0.99\linewidth]{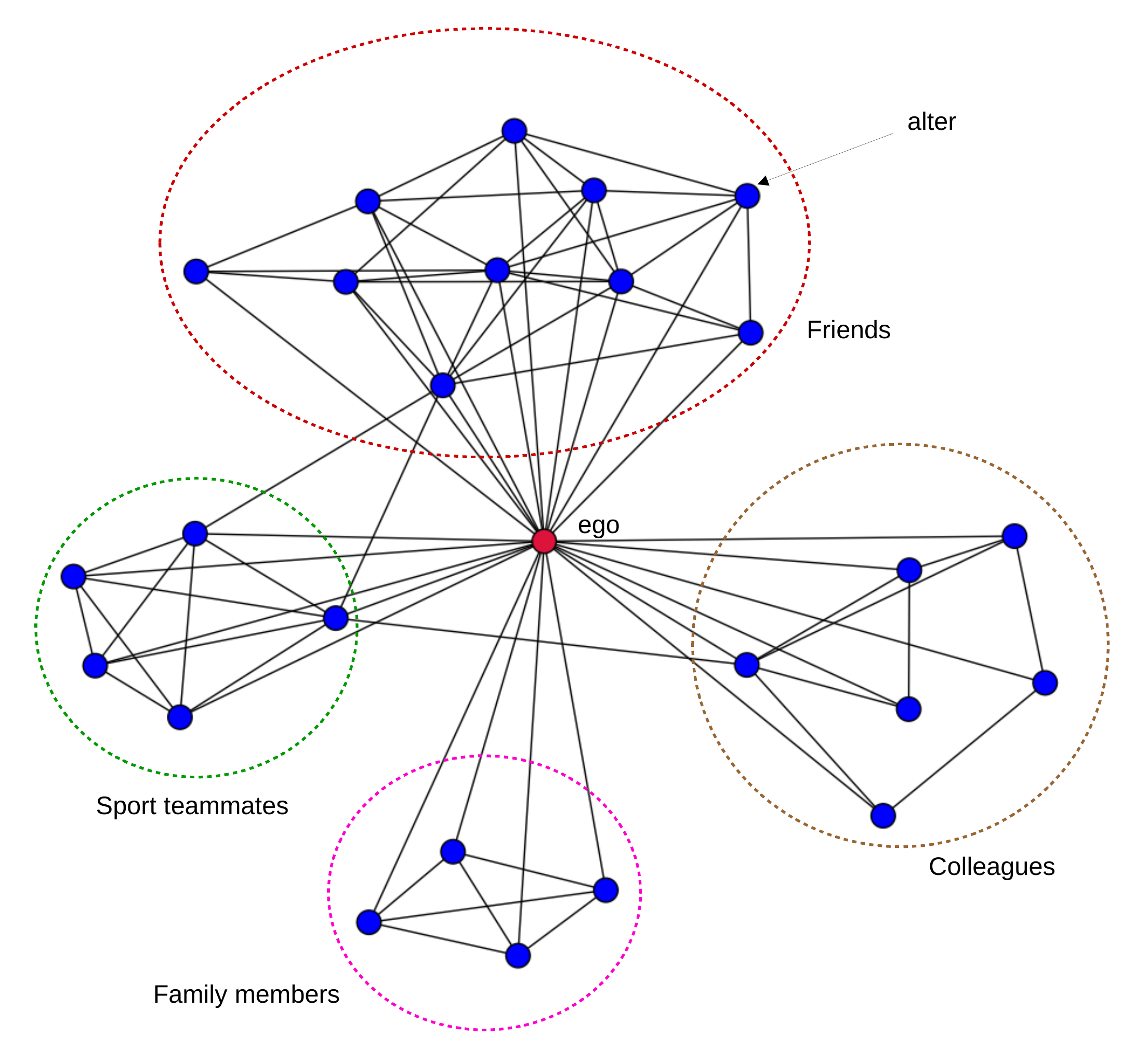}
\caption{An ego-network with four social circles}
\label{fig_1}
\end{figure}

With the exponential growth of social networks, extracting features
for nodes and detecting distinct neighborhood patterns become
increasingly impractical for the full network. One effective way to
succinctly describe certain aspects of large networks is breaking up
the network into smaller sub-networks \cite{ref1}. This is accomplished
by considering certain node or subgraph level locality statistics
specified on local regions of a network. These local regions are
defined as neighborhoods around a focal node (called ego). Therefore, ego-networks are social networks made up of an
ego along with all the social ties he has with
other people (called alters). Usually an ego categorizes his alters
into different groups (called social circles) such as family members,
friends, colleagues, etc. \autoref{fig_1} shows an exemplary
ego-network.

Ego-networks are an important subject of investigation in anthropology,
as several fundamental properties of social relationships can
be characterised by studying them. In particular, it has been
shown that neighborhoods around egos can exhibit different patterns. Based on prototypes of interactions between alters, prototypical neighborhood trends around egos can be dense, complete, star, etc. \cite{ref27}. Therefore, finding vector representations which convey the local neighborhood structure of egos is very helpful for ego-network analysis. These vector representations can easily exploited by statistical models for tasks such as social circle detection and prediction. Indeed, the local neighborhood analysis of nodes can reveal patterns and features of the network which are concealed
when only the global analysis is considered \cite{ref24}.

There has been several studies to learn vector representations for nodes based on global features. For instance, DeepWalk
\cite{ref9} learns global representations for nodes in the social graph utilizing deep learning techniques. In DeepWalk, first, nodes
are sampled from the underlying network to turn a network into a ordered sequence
of nodes same as way a document is an ordered sequence of words. Then, the Skip-gram
model \cite{ref7} is applied to learn feature representations for nodes by
optimizing a neighborhood preserving likelihood objective. Similarly, node2vec \cite{ref25} learns a mapping of nodes to a low-dimensional space of features which preserves the flexible notion of nodes' neighborhoods. Node2vec samples nodes using Breadth First Search or Depth First Search strategies. However, applying node2vec local sampling over ego-networks can not capture different structures since it exceeds the ego neighborhood.

Since an ego-network consists of certain number of alters, doing random walk over an ego-network can build an artificial paragraph. On the other hand, Paragraph Vector \cite{ref10} is an unsupervised framework which learns continuous distributed vector representations for pieces of texts. In this paper, we exploit the Paragraph Vector to learn neighborhood structures of egos in the social graph. Therefore, we investigate the interplay of global and local representations and
make the following contributions.
\begin{itemize}
\item We introduce local vector representations for nodes in
  ego-networks to complement the global representations for capturing
  the neighborhood structure; learning relations in a small
  neighborhood instead of relations in the entire
  graph. (\autoref{sec:Global-Local})
\item We apply global and local feature learning to the circle
  prediction problem. (\autoref{sec:circle-prediction})
\item We replace global representations by local representations to
  improve the performance. (\autoref{sec:circle-prediction})
\end{itemize}

The remainder of the paper is organized as follows. In
\autoref{sec:Global-Local}, we elaborate on global and local feature
learning of nodes in ego-networks. In \autoref{sec:circle-prediction},
we describe the problem of circle prediction and our approach. In
\autoref{sec:experiments}, we evaluate our approach on three datasets
from real-world social networks. We conclude in
\autoref{sec:conclusion}.

\section{Global and Local Feature Learning}
\label{sec:Global-Local}

An undirected graph is denoted by $G = (V, E)$, where $V$ is a set of
$n$ nodes and $E \subseteq V \times V$ is a set of edges. Furthermore,
let $G$ contain $m$ egos $\{u_1, u_2,\dots,u_m \} \in U \subseteq
V$. For an ego $u \in U$, we have the ego-network $G(u)$ as sub-graph
$\hat{G} = (\hat{V} ,\hat{E})$ of $G$, where $\hat{V}$ is the
neighborhood of $u$ and $\hat{E}$ is the intersection $E$ with
$\hat{V} \times \hat{V}$. We call a node in $\hat{V}$ an alter for $u$
and denote the set of alters of $u$ by $A_{u}$. Sets of alters for
different egos may overlap.

In this section, we apply the techniques which have been used to model
sentences and paragraphs of natural languages to model community
structure in networks. Therefore, we capture information on the global
and local network topology as follows:

\subsection{$\glo$: Learning global representation for each node}

According to DeepWalk, global feature learning consists of two main
components; first a random walk generator and second an update
procedure. Assume $\{v_1, v_2,\dots,v_n \} \in V$ are all nodes in the
graph $G$ , the idea is doing random walks started from every single
node. Then, having sequences of nodes such as
$ v_1 , v_2 ,\dots , v_{t−1}, v_t, \dots ,v_n $ with a context length
$c$, we update the representations to maximize the average log
probability:
\begin{equation}
 \sum_{t=1}^n \log P(v_t | v_{t+c} ,\dots, v_{t−c} ).
\end{equation}
Therefore, we have a mapping function $\glo \colon  V \to \RR^{|V| \times d}$, where $d$ is the embedding size.
\subsection{$\loc$: Learning local representation for each ego}
Inspired by Paragraph Vector, we learn a vector representation for every ego $u_i \in U$. Given ego $u_i$, first, we do random walks on $G(u_i)$ to compose an artificial paragraph which is called an ego-walk. This means an ego-walk is a stream of short random walks started at every $v_i \in A_{u_i} \cup \{u_i\} $. Then, having the ego-walk $ v_1, v_2, v_3 ,\dots,v_t,\dots, v_l $ for ego $u_i$, we aim to update the representations in order to maximize the average log probability:
\begin{equation}
\sum_{t=1}^l \log P(v_t|u_i, v_{t+c},\dots, v_{t-c})
\end{equation}
Where $l$ is the length of the ego-walk with $l<n$, and $c$ is the context length.
Therefore, we introduce a mapping function $\loc \colon U \to \RR^{|U| \times d}$, where $d$ is the embedding size.

\begin{figure}[H]
\centering
\includegraphics[width=0.88\linewidth]{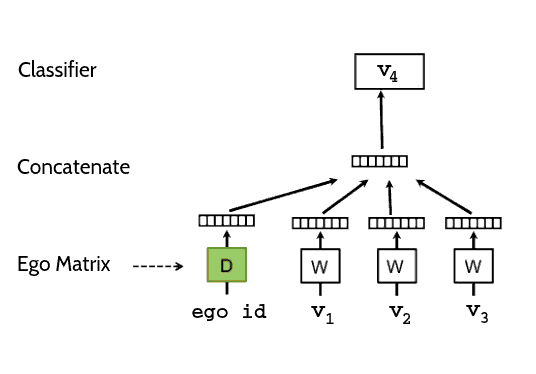}
\caption{A technique for learning an ego vector. The concatenation of
  an ego vector with the context of three alters is used to predict
  the fourth alter.}
\label{fig_2}
\end{figure}

In our technique (see \autoref{fig_2}), every
ego is mapped to a unique vector, represented by a
column in ego matrix D and every alter is also mapped to a
unique vector, represented by a column in matrix W . The
ego vector and alter vectors are concatenated to predict the next alter in a context.

\section{circle prediction}
\label{sec:circle-prediction}

In online social networks, users need to organize their
personal social networks to cope with the information overload generated by
their friends \cite{ref1}. However, this
manual process is laborious, error-prone and inadaptable to changes.
It is meaningful and essential to study how to automatically organize user's friends into social
circles when they are added to the network. These organized social circles could help solve many practical problems. For example, it can preserve user's privacy by showing updates and information only to some friends belong to the specific circles allowed by the user. It also can help a user who wants to read the latest news
from his colleagues instead of scrolling through all the news
from other users.

However, most of current social circle identification methods
\cite{ref1,ref2,ref3,ref4,ref5,ref11,ref12,ref13}
are unsupervised learning methods which lacks emphasis on dataset
quality and they could not predict well when there is a missing value
in the query. The main supervised approach is proposed by McAuley \&
Leskovec \cite{ref1} which trained a binary classifier for each
circle. Their probabilistic model discriminates members from
nonmembers based on node features. Node features are the information
from both network topological structure and users' profiles. Although
their model deals with weak supervision to predict the circle for a
new alter, it fails to refit the model for every new alter that is
added to the network.  In this section, we study the problem of social
circle prediction exploiting the global and local neighborhood
structures.
\subsection{Approach}
We formulate the problem of circle prediction as a classification task on a new added alter into the graph. We thus leverage the topological structure of the alter and also his profile information. Indeed, Alters' and egos' profile information help with the circle prediction task. For example, if the ego and the alter both go to the same university, probably this alter belongs to the university friends circle. Therefore, we add the common profile features vector between an ego and its alter to the topological representations to perform a more accurate circle prediction. More formally, we denote $i^{\text{th}}$ profile feature of the alter $v$ as $v.feat_i$, and the ego $u$ as $u.feat_i$. Given ego $u \in U$, and alter $v \in V$, we formulate the ego and alter profile similarity as $\simi(u,v) = (b_{1}, \dots, b_{f})$, where

\begin{equation}
  b_{i}=
     \begin{cases}
     1 & \text{if $ u.\feat_{i} = v.\feat_{i}$}, \\
     0 & \text{otherwise}.
     \end{cases}
\end{equation}
Therefore, for each pair of ego and alter, we have the binary vector $\simi(u,v) \in \RR^{f}$ where $f$ is the number of profile features.
\subsection{Classifier}
Since some alters are the member of several circles, we need to use a multi-label classifier. Neural network classifiers have the ability to detect all possible interactions between predictor variables. Furthermore, they need less formal statistical training to develop \cite{ref14}. In particular, feed-forward neural networks are appropriate for modeling relationships between a set of input variables and one or more output variables. In fact, they are suitable for any functional mapping problem where we want to know how a number of input variables affect the output variable \cite{ref18}. We thus define our classifier as a multi-layer feed-forward neural network with the following possible input layers:
\begin{itemize}
\item Where the input layer is the concatenation ($\oplus$) of the global and local representations:

\begin{itemize}
\item \textbf{locglo:} $\loc(u) \oplus \glo(v)$
\item \textbf{gloglo:} $\glo(u) \oplus \glo(v) $
\item \textbf{locgloglo:} $\loc(u) \oplus \glo(u) \oplus glo(v) $
\end{itemize}

\item Where the input layer is the the concatenation of global representation, local representation and the profile similarity vector:
\begin{itemize}
\item \textbf{locglosim:} $\loc(u) \oplus \glo(v \oplus \simi(u,v)$
\item \textbf{gloglosim:} $\glo(u) \oplus \glo(v) \oplus \simi(u,v)$
\item \textbf{locgloglosim:} $\loc(u) \oplus \glo(u) \oplus \glo(v) \oplus \simi(u,v) $
\end{itemize}

\end{itemize}

Overall, the architecture of our classifier is described as follows:
\begin{itemize}
\item \textbf{Input layer:} It can be one of six possible inputs which were described above.
\item \textbf{Hidden layer:} We have a hidden layer with ReLU activation unit \cite{ref21}.
\item \textbf{Output layer:} The output layer has $|\mathcal{Y}|$ units the same as the number of social circles in the graph with softmax activation function \cite{ref22}.
\item \textbf{Optimizer:} we used RMSprop which is an adaptive learning rate method that has found much success in practice
\cite{ref19}. RMSprop divides	the	learning	rate	for	a
weight	by	a	running	average	of	the
magnitudes	of	recent	gradients	for	that
weight.
%
%
%
\end{itemize}

\section{Experiments}
\label{sec:experiments}
In this section, we first provide an overview of the datasets that we
used in the experiments. We then present an experimental
analysis of the proposed approach.

\subsection{Datasets}
Since our approach is supervised, we require labeled ground-truth data in order to evaluate its performance. We obtained ego-networks and
ground-truth from three major social networking sites: Facebook, Google+, and Twitter available from the University of Stanford
\cite{ref1}. \autoref{tab:dataset} describes the details of the datasets we used in our experiments.

\begin{table}[H]
\centering
  \caption{Statistics of Social Network Datasets}
  \label{tab:dataset}
\begin{tabular}{lcccc}
  \toprule
  \head{}  & \head{}          & \head{Facebook} & \head{Twitter} & \head{Google+} \\
  \midrule
  nodes    & $\vert V \vert $ & 4,039           & 81,306         & 107,614        \\
  edges    & $\vert E \vert $ & 8,8234          & 1,768,149      & 13,673,453     \\
  egos     & $|U|$            & 10              & 973            & 132            \\
   circles & $|\mathcal{Y}|$  & 46              & 100            & 468            \\
  features & $f$              & 576             & 2271           & 4122           \\
  \bottomrule
\end{tabular}
\end{table}

The number of circles refers to the number of different social circles such as family members, highschool, sport, colleagues, etc.

\subsection{Experimental setup}

In order to learn global representations for nodes in
Facebook, Google+, and Twitter graphs, we first do random walks to compose three artificial corpus. We then apply word2vec of gensim \cite{ref23} which is an implementation for the Skip-gram model on our artificial corpuses. We set the embedding size
$d=300$ \cite{ref26}, and the context length $c=2$. Therefore, word2vec scans over the nodes, and for each
node it tries to embed it such that the node's features can predict nearby nodes. The node feature representations are learned by optimizing the likelihood objective using SGD with negative sampling \cite{ref8}.

Similarly, we set the embedding size
$d=300$ and $c=2$ to learn local representations for egos in these social graphs. First, we generate ego-walks doing random walks on each ego-network separately. For example, for the Facebook graph with $10$ egos, we have a corpus with $10$ ego-walks. Then, we apply doc2vec of gensim \cite{ref23} which modifies the word2vec algorithm to learn continuous representations for paragraphs on our artificial corpuses. Therefore, every ego is represented by a vector which holds the semantics of his neighborhood structure.

To obtain common features for each pair of ego and alter, we select the first $500$ features of their profiles include birthday, education, gender, hometown, languages, location, work along with their sub-branches. We then compare the ego's features to his alters' features one by one to generate a binary feature vector. This vector will be concatenated to the topological structure vectors as input of the classifier.

We create feature matrices $X_{locglo} \in \RR^{2d}$ and $X_{gloglo} \in \RR^{2d}$ by concatenation of local and global vectors where $d= 300$. We also create two other feature matrices $X_{locglosim}\in \RR^{2d+f}$ and $X_{gloglosim}\in \RR^{2d+f}$ considering common profile feature vectors where $f=500$. The same manner we have $X_{locgloglo} \in \RR^{3d}$ and $X_{locgloglosim}\in \RR^{3d+f} $.

Regarding to the ground-truth matrix, we have circle labels for each alter available in the dataset. We need to convert the multi-label ground-truth to the binary form which is more suitable for the classification algorithm.

We finally perform the classification task considering different
inputs $X_{loc}$, $X_{glo}$, $X_{locsim}$, $X_{glosim}$, $X_{locglo}$ and $X_{locglosim}$ to compare the prediction results.
In the multi-label classification setting, every alter is assigned one
or more labels from a finite set $\mathcal{Y}$. During the training phase, we
observe a $70\%$ of alters and all their labels. The task is to
predict the labels for the remaining $30\%$ alters. The batch size of
the stochastic gradient descent is set to $32$ for Facebook and $64$
for both Google+ and Twitter since they have bigger graphs. We consider the
learning rate $\epsilon=0.001$ for RMSprop optimizer over $50$
iterations. We use $K$-fold cross-validation approach for estimating test error. The idea is to randomly divide the data into $K$ equal-sized. We leave out part $k$, fit the model to the other
$K-1$ parts (combined), and then obtain predictions for
the left-out $k^{\text{th}}$ part. This is done in turn for each part $k = 1, 2, \cdots K$, and then
the results are combined. We set $K=10$ in our experiments.

\subsection{Results}
We classify the alters of Facebook, Google+, and Twitter graphs into respective social circles and report the average
performance in terms of $F_1$-score. To compute the $F_1$-score we
follow evaluation metrics was described as \cite{ref1} with 10-fold cross validation. \autoref{tab:fscore1}
shows the average performance of the classifier.
As can be seen, replacing global representation
with local improved the performance of the circle prediction. Moreover, considering the profile similarity between
ego and alter affected on the performance of the classifier.
However, adding the global representations of egos to the
input did not improve the performance.

\begin{table}[H]
\centering
\caption{Performance ($F_1$-score) of different embeddings for circle
  prediction on three dataset. Standard deviation is less than $0.02$ for all experiments.}
\label{tab:fscore1}
\begin{tabular}{p{2 cm}ccc}
  \toprule
  \head{Approach} & \head{Facebook} & \head{Twitter} &  \head{Google+} \\
  \midrule
gloglo & 0.37 & 0.46 & 0.49 \\
locglo & 0.42  & 0.50 & 0.52 \\
locgloglo & 0.37 & 0.44 & 0.48 \\
\midrule
gloglosim & 0.40 & 0.49 & 0.51 \\
locglosim  & 0.45 & 0.53 & 0.55 \\
locgloglosim & 0.38 & 0.46 & 0.47 \\
\midrule
$\Phi^{1}$, McAuley~\& Leskovec \cite{ref1} & 0.38 & 0.54 & 0.59 \\
  \bottomrule
\end{tabular}
\end{table}

\section{Conclusion}
\label{sec:conclusion}

We described a technique for ego-network analysis based
on the concept of local network neighborhoods. We applied new advancements of language modeling to learn latent social representations for egos. This allows
analysis on large social networks and can reveal aspects of
neighborhood structure that cannot be ascertained in a
global network analysis. We provided an example of social circle prediction on different social graphs displaying the ability of our
approach to capture local neighborhood structure. As a future work, we tend to study how the local representations can improve the other graph analysis tasks (e.g. link prediction, shortest path, etc).

\end{document}